\newcounter{myctr}
\def\myitem{\refstepcounter{myctr}\bibfont\noindent\ifnum\themyctr>9\else\phantom{0}\fi\hangindent17pt\themyctr.\enskip}

\documentclass{ws-ijqi}

\begin{document}

\markboth{V.I. Man'ko, L.A. Markovich}
{NEW MINKOWSKI TYPE INEQUALITIES $\ldots$}

\catchline{}{}{}{}{}

\title{NEW MINKOWSKI TYPE INEQUALITIES AND ENTROPIC INEQUALITIES FOR QUANTUM STATES OF QUDITS}

\author{V.I. MAN'KO\footnote{manko@sci.lebedev.ru}}

\address{P.N. Lebedev Physical Institute, Russian Academy of Sciences,\\
Leninskii Prospect 53, Moscow 119991, Russia}

\author{L.A. MARKOVICH\footnote{kimo1@mail.ru}}

\address{Moscow Institute of Physics and Technology, State University, Moscow, Russia}

\maketitle

\begin{history}
\received{Day Month Year}
\revised{Day Month Year}
\end{history}

\begin{abstract}
The two-parameter Minkowski like inequality written for composite quantum system state is obtained for arbitrary Hermitian nonnegative matrix
with trace equal to unity. The inequality can be used as entropic and information inequality for density matrix of noncomposite finite quantum system, e.g.,
for a single qudit state. The analogs of strong subadditivity condition for the single qudit is discussed in context of obtained Minkowski like inequality.
\end{abstract}


\section{Introduction}	

There exist inequalities called  Minkowski type trace inequalities \cite{Lieb1,Lieb2} for density matrices of bipartite system. The inequalities reflect some properties
of the quantum correlations between subsystem degrees of freedom  in the composite quantum system. The inequalities are  tracial analog of known  Minkowski's inequality for multiple integrals \cite{Hardy}. The Minkowski type inequalities have two forms. The first form corresponds to density matrix inequalities for bipartite finite quantum system depending on a single parameter \cite{Lieb1}. Another density matrix inequality for bipartite quantum system depends on two parameters \cite{Lieb2}. In Ref.~\refcite{Cher} the one-parameter Minkowski type inequality
was generalized for the case of a single qudit state which does not contain subsystems.
\par The aim of our article is to obtain the generalization of two-parameter Minkowski type  inequality known for bipartite systems to the case of the single qudit system, in particular for the qutrit and qudit with $j=3/2$. We apply the tool obtained in Ref.~\refcite{Chernega,Chernega:14,OlgaMankoarxiv,Manko:2014} to use a bijective map of the Hilbert space of a single quantum system onto the Hilbert space $\mathcal{H}^1\otimes\mathcal{H}^2$ of bipartite quantum system. The tool is similar to the method used to consider the Laplace matrices \cite{Braunstein,Bose}. We shall study the one- and two-parameter Minkowski like inequalities on examples of Werner state \cite{Werner} and $X$-state  \cite{Hedemann}. We extend the notion of $X$-state to the case of qudit with $j=3/2$. Also
we apply the method of embedding  the qutrit and qudit $j=3/2$ spaces of states into the Hilbert spaces of higher dimensions. The new Minkowki type inequality for qutrit and qudit with $j=3/2$ are written in an explicit form of matrix inequalities.
\par The article is organized as follows. In Sec.~\ref{sec:2} the one-parameter Minkowski type trace inequality is written for the qudit with $j=3/2$. The simulation on examples of Werner and $X$-state density matrices is presented. In Sec.~\ref{sec:3} we remind characteristics of the degree of quantum correlations in the two-qubit system introduced in Ref.~\refcite{Markovich2:2014} on the example of the Werner density matrix for two-qubit system. In Sec.~\ref{sec:4} the two-parameter Minkowski type inequality  for the qutrit and qudit systems is obtained. The results are demonstrated on the example of the Werner density matrix for the qudit with spin $j=3/2$. The conclusion and perspectives are given
in Sec.~\ref{sec:5}.

\section{The One-Parameter Minkowski Type Trace Inequality}\label{sec:2}

Let us define a quantum system on the Hilbert space. The density matrix of the system state in a four-dimensional Hilbert space $\mathcal{H}$ is the following
\begin{eqnarray}\rho={\left(
                                 \begin{array}{cccc}
                                   \rho_{11}& \rho_{12}& \rho_{13}& \rho_{14}\\
                                   \rho_{21}& \rho_{22}& \rho_{23}& \rho_{24}\\
                                   \rho_{31}& \rho_{32}& \rho_{33}& \rho_{34}\\
                                   \rho_{41}& \rho_{42}& \rho_{43}& \rho_{44}\\
                                 \end{array}
                               \right)}\,. \label{1}
                               \end{eqnarray}
The latter matrix has standard properties of the density matrix, namely,  $\rho=\rho^{\dagger}$ and $Tr\rho=1$  hold and its eigenvalues are nonnegative.
We can partition the matrix \eqref{1}  into four $2\times2$ blocks
\begin{eqnarray*}a_{11}&=&\left(
                            \begin{array}{cc}
                              \rho_{11} & \rho_{12}  \\
                              \rho_{21}  & \rho_{22}  \\
                            \end{array}
                          \right),\quad a_{12}=\left(
                            \begin{array}{cc}
                              \rho_{13} & \rho_{14}  \\
                              \rho_{23}  & \rho_{24}  \\
                            \end{array}
                          \right),\quad
                          a_{21}=\left(
                            \begin{array}{cc}
                              \rho_{31} & \rho_{32}  \\
                              \rho_{41}  & \rho_{42}  \\
                            \end{array}
                          \right),\quad a_{22}=\left(
                            \begin{array}{cc}
                              \rho_{33} & \rho_{34}  \\
                              \rho_{43}  & \rho_{44}  \\
                            \end{array}
                          \right).
\end{eqnarray*}
Hence, the partitioned matrix can then be written as
\begin{eqnarray*}\rho&=&\left(
                            \begin{array}{cc}
                              a_{11} & a_{12}  \\
                             a_{21}  & a_{22}  \\
                            \end{array}
                          \right).
\end{eqnarray*}
Let us introduce the following notation
\begin{eqnarray*}&&\left(
                            \begin{array}{cc}
                              a_{11} & a_{12}  \\
                             a_{21}  & a_{22}  \\
                            \end{array}
                          \right)^p\equiv\left(
                            \begin{array}{cc}
                              a_{11}(p) & a_{12}(p)  \\
                             a_{21}(p)  & a_{22}(p)  \\
                            \end{array}
                          \right),
\end{eqnarray*}
where $p\geq1$ is a given real number. The $\{a_{ij}(p)\}$ are new block matrices which depend on the parameter $p$. The following  inequality 
\begin{eqnarray}{\left(Tr\left(a_{11}+a_{22}\right)^p\right)^{\frac{1}{p}}\leq
 Tr\left(
                            \begin{array}{cc}
                              Tr a_{11}(p) & Tr a_{12}(p)  \\
                             Tr a_{21}(p)  &  Tra_{22}(p)  \\
                            \end{array}
                          \right)^{\frac{1}{p}}}\, \label{13}
\end{eqnarray}
holds.
 The inequality reverses in case $0\leq p<1$. This inequality is one-parameter Minkowki type inequality for the qudit with $j=3/2$.
\par
\begin{figure}[htbp]
\centerline{\psfig{file=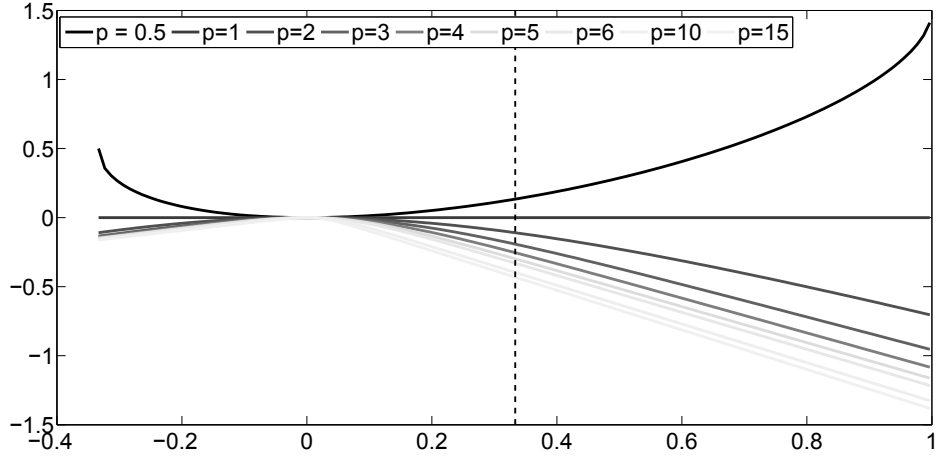, width=15cm}}
\vspace*{8pt}
\caption{The $\Delta \mathcal{M}(p)$ (Y-axes) for $p=\{1,2,3,4,5,6,10,15\}$ against the parameter $r$ (X-axes), where the dashed line marks $r=1/3$.\label{fig:1}}
\end{figure}

Let us denote the  residual of the left and right sides in \eqref{13} as $\Delta \mathcal{M}(p)$.
For example, if the parameter $p$ is equal to $2$ then the latter residual can be represented as
\begin{eqnarray*}\Delta \mathcal{M}(2)&=&\rho_{11}\rho_{33}+ \rho_{22}\rho_{44} - \rho_{13}\rho_{31} + \rho_{12}\rho_{43}\\
 &-& \rho_{14}\rho_{41} + \rho_{21}\rho_{34} - \rho_{23}\rho_{32}  - \rho_{24}\rho_{42}\leq0.
\end{eqnarray*}
As an example of the qudit state with $j=3/2$  one can take the Werner state \cite{Werner}. It is determined by density matrix  of the form
\begin{eqnarray}\rho^{W}(r)={\left(
                     \begin{array}{cccc}
                       \frac{1+r}{4} & 0 & 0 & \frac{r}{2}\\
                       0 & \frac{1-r}{4}& 0 & 0 \\
                       0 & 0 & \frac{1-r}{4} & 0 \\
                       \frac{r}{2} & 0 & 0 & \frac{1+r}{4} \\
                     \end{array}
                   \right)}\,, \label{2}
\end{eqnarray}
where  $-1/3\leq r\leq1$. The parameter domain $1/3< r\leq1$ corresponds to the entangled state.
Hence, for the $p=2$ the inequality \eqref{13} is determined by
\begin{eqnarray*}&&\Delta \mathcal{M}(p)=1-3r^2\leq0
\end{eqnarray*}
which leads to $1/3< r$.
The  $\Delta \mathcal{M}(p)$ for the Werner state is shown in Figure~\ref{fig:1} for various values of the parameter $p$.

\par If $\rho_{12}=\rho_{13}=\rho_{21}=\rho_{31}=\rho_{24}=\rho_{34}=\rho_{42}=\rho_{43}=0$ holds than the density matrix \eqref{1} has the form of the  $X$-state density matrix
                               \begin{eqnarray}\label{14}
\rho^{X}&=&\left(
                     \begin{array}{cccc}
                       \rho_{11} & 0 & 0 & \rho_{14}\\
                       0 & \rho_{22}& \rho_{23} & 0 \\
                       0 & \rho_{32} & \rho_{33} & 0 \\
                       \rho_{41} & 0 & 0 & \rho_{44} \\
                     \end{array}
                   \right)=\left(
                     \begin{array}{cccc}
                       \rho_{11} & 0 & 0 & \rho_{14}\\
                       0 & \rho_{22}& \rho_{23} & 0 \\
                       0 & \rho_{23}^{\ast} & \rho_{33} & 0 \\
                      \rho_{14}^{\ast} & 0 & 0 & \rho_{44} \\
                     \end{array}
                   \right),
\end{eqnarray}
where $\rho_{11},\rho_{22},\rho_{33},\rho_{44}$ are real positive and $\rho_{23},\rho_{14}$ are complex quantities. Thus, it is possible to write
the Minkowki type inequalities \eqref{13} for the $X$-state density matrix of the qudit state with spin $j = 3/2$ \cite{Markovich3:2014}.
For example, for $p=2$ the inequality \eqref{13} has the following form
\begin{eqnarray*}&&\rho_{11}\rho_{33} + \rho_{22}\rho_{44}- |\rho_{14}|^2 - |\rho_{23}|^2 \leq 0.
\end{eqnarray*}

\section{Inequalities for Quantum von Neumann and Tomographic
Mutual Information}\label{sec:3}

The tomographic probability representation of spin (qudit) states was introduced in Ref.~\refcite{Dodonov,OVManko} .
In this representation the qudit states with some density matrix $\rho$ are identified with the spin-tomogram which is the probability distribution function determined
by the density operator of the states. The spin tomogram is determined by
\begin{eqnarray*}\omega(m_1,m_2,\overline{n}_1,\overline{n}_2)&=&\langle m_1,m_2|U\cdot\rho\cdot U^{\dagger}|m_1,m_2\rangle,
\end{eqnarray*}
where  $m_{1,2}=-j,-j+1,\ldots,j$, $j= 0,1/2,1\ldots$ are spin projections. The rotation matrix $U$ is defined as the direct product of two matrices of irreducible representations of $SU(2)$ - group
\begin{eqnarray*}
U&=&U(\theta_1,\varphi_1,\psi_1)\otimes U(\theta_2,\varphi_2,\psi_2),\\
U(\theta_k,\varphi_k,\psi_k)&=& \left(
                       \begin{array}{cc}
                         \cos\frac{\theta_k}{2} e^{\frac{i(\varphi_k+\psi_k)}{2}}& \sin\frac{\theta_k}{2} e^{\frac{i(\varphi_k-\psi_k)}{2}} \\
                         -\sin\frac{\theta_k}{2}e^{\frac{i(\psi_k-\varphi_k)}{2}} & \cos\frac{\theta_k}{2} e^{\frac{-i(\varphi_k+\psi_k)}{2}}\\
                       \end{array}
                     \right).
\end{eqnarray*}
\par The Werner state of two qubits can be taken as the quantum state which can be either separable or entangled depending on the parameter values of
 its density  matrix. The eigenvalues of the density matrix  \eqref{2} are
\begin{eqnarray*}&&\lambda_1 = \frac{1+3r}{4}, \quad \lambda_{2,3,4} = \frac{1-r}{4}.
\end{eqnarray*}
Hence, the reduced density matrices of the first and the second qubits are the following
\begin{eqnarray*}\rho_1&=&\left(
                            \begin{array}{cc}
                              \frac{1}{2} & 0 \\
                              0 & \frac{1}{2} \\
                            \end{array}
                          \right),
                          \rho_2=\left(
                            \begin{array}{cc}
                              \frac{1}{2} & 0 \\
                              0 & \frac{1}{2} \\
                            \end{array}
                          \right).
\end{eqnarray*}
The von Neumann entropies of both qubit states and the entropy of the whole system are
\begin{eqnarray*}S_1 &=& -Tr\rho_1\ln\rho_1 = \ln2, \quad S_2 =-Tr\rho_2\ln\rho_2= \ln2,\\ \nonumber
S_{12} &=&-Tr\rho \ln\rho
=-\frac{1+3r}{4}\ln\left(\frac{1+3r}{4}\right)-3\frac{1-r}{4}\ln\left(\frac{1-r}{4}\right).
\end{eqnarray*}
The quantum information is defined as
\begin{eqnarray*}I_q&=&S_1 +S_2-S_{12}
\end{eqnarray*}
and it satisfies the inequality $I_q\geq 0$.
\par The diagonal matrix elements of  $U\cdot\rho\cdot U^{\dagger}$ are
\begin{eqnarray*}&&\omega(\uparrow,\uparrow)=\omega(\downarrow,\downarrow)
=\frac{1}{4}\left(r\left(\cos\theta_1\cos\theta_2 + \cos(\psi_1 + \psi_2)\sin\theta_1\sin\theta_2\right) + 1\right),\\\nonumber
&&\omega(\uparrow,\downarrow)=\omega(\downarrow,\uparrow)=\frac{1}{4}\left(1 - r\left(\cos(\psi_1 + \psi_2)\sin\theta_1\sin\theta_2 + \cos\theta_1\cos\theta_2\right)\right).
\end{eqnarray*}
The $\omega(\uparrow,\uparrow)\equiv\omega\left(+\frac{1}{2},+\frac{1}{2},\overline{n}_1,\overline{n}_2\right)$ denotes the  tomographic probability.
The  trace of the rotated density matrix satisfies the normalization condition
$Tr\left(U\cdot\rho\cdot U^{\dagger}\right)=1$. Marginal distributions corresponding to the first and the second qubit density matrices are
\begin{eqnarray*}W_1(\uparrow,\overline{n}_1)&=&\omega(\uparrow,\uparrow)+\omega(\uparrow,\downarrow),\quad
W_1(\downarrow,\overline{n}_1)=\omega(\downarrow,\uparrow)+\omega(\downarrow,\downarrow),\\
W_2(\uparrow,\overline{n}_2)&=&\omega(\uparrow,\uparrow)+\omega(\downarrow,\uparrow),\quad
W_2(\downarrow,\overline{n}_2)=\omega(\uparrow,\downarrow)+\omega(\downarrow,\downarrow).
\end{eqnarray*}
By definition of Shannon entropy we can construct the tomographic entropies of the qubit subsystems as
\begin{eqnarray*}H_1 &=& -W_1(\uparrow,\overline{n}_1)\ln W_1(\uparrow,\overline{n}_1)-W_1(\downarrow,\overline{n}_1)\ln W_1(\downarrow,\overline{n}_1)=\ln2,\\ \nonumber
H_2&=& -W_2(\uparrow,\overline{n}_2)\ln W_2(\uparrow,\overline{n}_2)-W_2(\downarrow,\overline{n}_2)\ln W_2(\downarrow,\overline{n}_2)=\ln2.
\end{eqnarray*}
The tomographic Shannon entropy  of the bipartite system reads
\begin{eqnarray*}
H_{12}&=&-\omega(\uparrow,\uparrow)\ln\omega(\uparrow,\uparrow) - \omega(\uparrow,\downarrow)\ln\omega(\uparrow,\downarrow)-
\omega(\downarrow,\uparrow)\ln\omega(\downarrow,\uparrow) - \omega(\downarrow,\downarrow)\ln\omega(\downarrow,\downarrow).
\end{eqnarray*}
We define the information by
\begin{eqnarray*}\label{5}I_t &=& \max\limits_{\psi_1,\psi_2,\theta_1,\theta_2}(H_1+H_2-H_{12}).
\end{eqnarray*}
It satisfies the inequality $I_t\geq0$.
The residual  of the
 quantum information $I_q$ and the maximum
of the unitary tomographic information $I_t$ is
\begin{eqnarray}\label{4}\triangle I&=&I_q -I_t \geq0.
\end{eqnarray}
The residual $\triangle I$ for the maximal entangled Werner state is
\begin{eqnarray*}&&\lim\limits_{r\rightarrow1}\triangle I= \ln2\approx0.693.
\end{eqnarray*}

\subsection{Comparison with negativity and concurrence}

\begin{figure}[htbp]
\centerline{\psfig{file=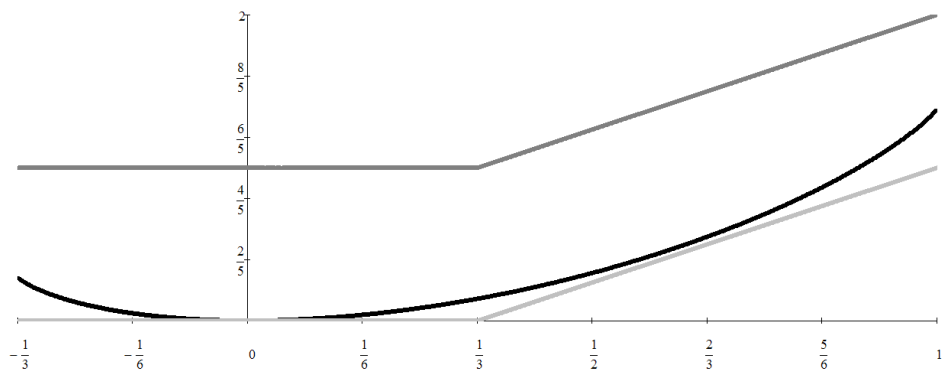, width=13cm}}
\vspace*{8pt}
\caption{The $\triangle I$ (black line), the concurrence $C(r)$ (light gray line), the negativity $N(r)$ (dark gray line) against $r$.\label{fig:4}}
\end{figure}

    For the Werner state \eqref{2} the negativity and the concurrence are determined by
    \begin{eqnarray*}N(r)&=&3\left|\frac{r+1}{4}\right|+\left|\frac{1-3r}{4}\right|>1,\\
    C(r)&=& \left\{
\begin{array}{ll}
\frac{3r-1}{2}, &   \mbox{if}\qquad \frac{1}{3}<r\leq1,
\\
0, & \mbox{if}\qquad -\frac{1}{3}\leq r\leq\frac{1}{3},
\end{array}
\right.
 \end{eqnarray*}
respectively. The comparison of the negativity, the concurrence and \eqref{4} is shown in Figure \ref{fig:4}.

\section{The Two-Parameter Minkowski Type Trace Inequality}\label{sec:4}

Analogously to Sec.~\ref{sec:2}, the two-parameter Minkowski type trace inequality can be introduced for the qudit system.
Let us define the qudit density matrix in the block form
\begin{eqnarray}\rho={\left(
                     \begin{array}{cccc}
                       a_{11} &a_{12} & \cdots & a_{1n}\\
                       a_{21} & a_{22}& \cdots & a_{2n} \\
                       \vdots & \vdots & \ddots & \vdots \\
                       a_{n1} & a_{n2} & \cdots & a_{nn} \\
                     \end{array}
                   \right)}\,, \label{3}
\end{eqnarray}
where $\rho=\rho^{\dagger}$, $Tr\rho=1$, $\rho\geq0$. For the given real numbers $p$ and $q$ we introduce the following notations
\begin{eqnarray*}&&\rho^p=\left(
                     \begin{array}{cccc}
                       a_{11}(p) &a_{12}(p) & \cdots & a_{1n}(p)\\
                       a_{21}(p) & a_{22}(p)& \cdots & a_{2n}(p) \\
                       \vdots & \vdots & \ddots & \vdots \\
                       a_{n1}(p) & a_{n2}(p) & \cdots & a_{nn}(p) \\
                     \end{array}
                   \right),\quad
                   \rho^q=\left(
                     \begin{array}{cccc}
                       a_{11}(q) &a_{12}(q) & \cdots & a_{1n}(q)\\
                       a_{21}(q) & a_{22}(q)& \cdots & a_{2n}(q) \\
                       \vdots & \vdots & \ddots & \vdots \\
                       a_{n1}(q) & a_{n2}(q) & \cdots & a_{nn}(q) \\
                     \end{array}
                   \right).
\end{eqnarray*}
The new Minkowski type trace inequality for the qudit system reads
\begin{eqnarray}{\left(Tr\left(\sum\limits_{i=1}^na_{ii}(q)\right)^{\frac{p}{q}}\right)^{\frac{1}{p}}\leq
\left(Tr\left(
                     \begin{array}{cccc}
                       Tr a_{11}(p) & Tr a_{12}(p) & \cdots & Tr a_{1n}(p)\\
                       Tr a_{21}(p) & Tr a_{22}(p)& \cdots & Tr a_{2n}(p) \\
                       \vdots & \vdots & \ddots & \vdots \\
                       Tr a_{n1}(p) & Tr a_{n2}(p) & \cdots & Tr a_{nn}(p) \\
                     \end{array}
                   \right)^{\frac{q}{p}}\right)^{\frac{1}{q}}}\,. \label{18}
\end{eqnarray}
The inequality holds for all $1\leq p\leq2$, otherwise it should be reversed. 
\par
\begin{figure}[htbp]
\centerline{\psfig{file=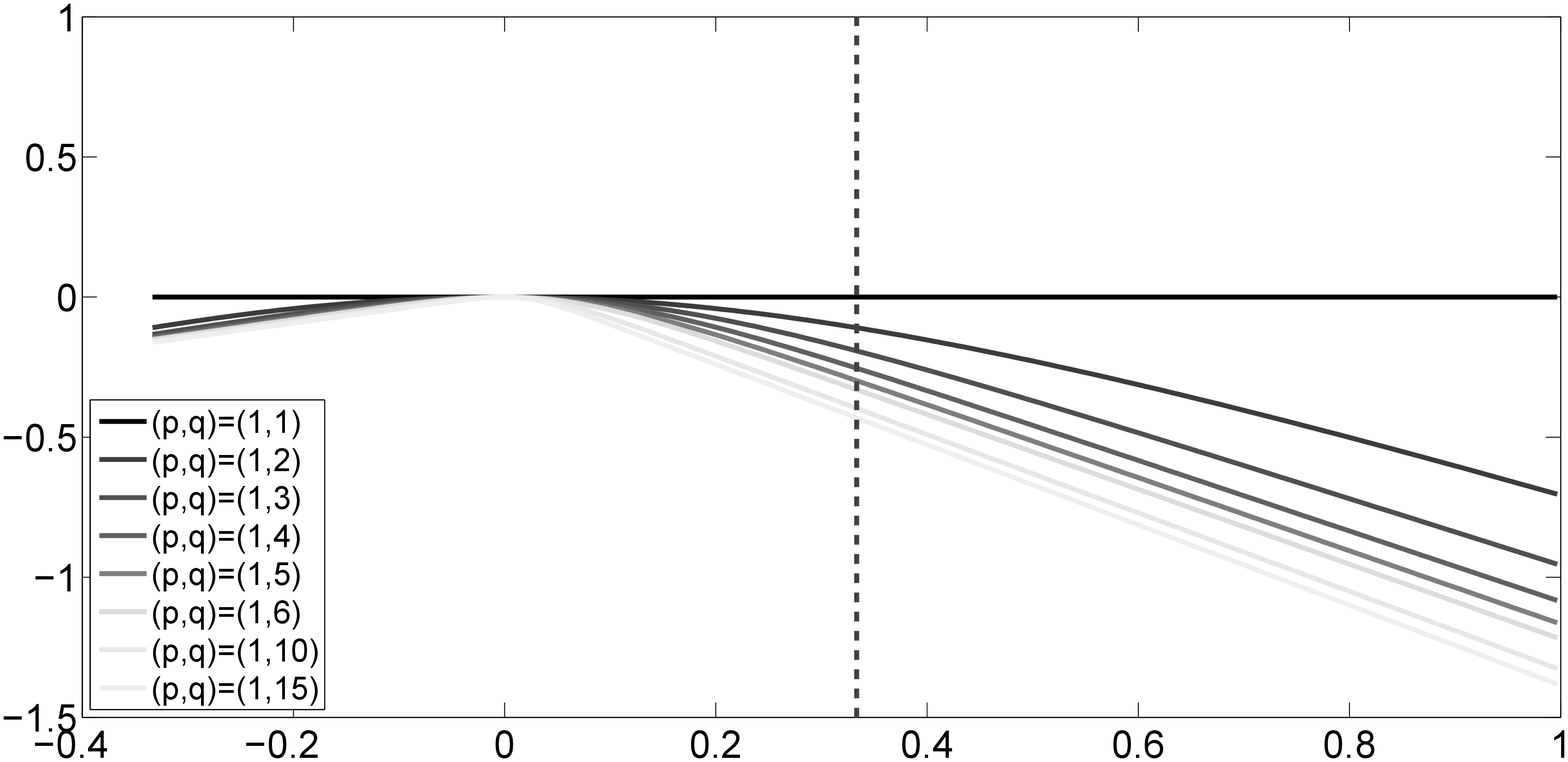, width=15cm}}
\vspace*{8pt}
\caption{The $\Delta \mathcal{M}(p,q)$ (Y-axes) for various pairs $(p,q)$ against $r$ (X-axes), where the dashed line marks $r=1/3$.\label{fig:2}}
\end{figure}

\begin{figure}[htbp]
\centerline{\psfig{file=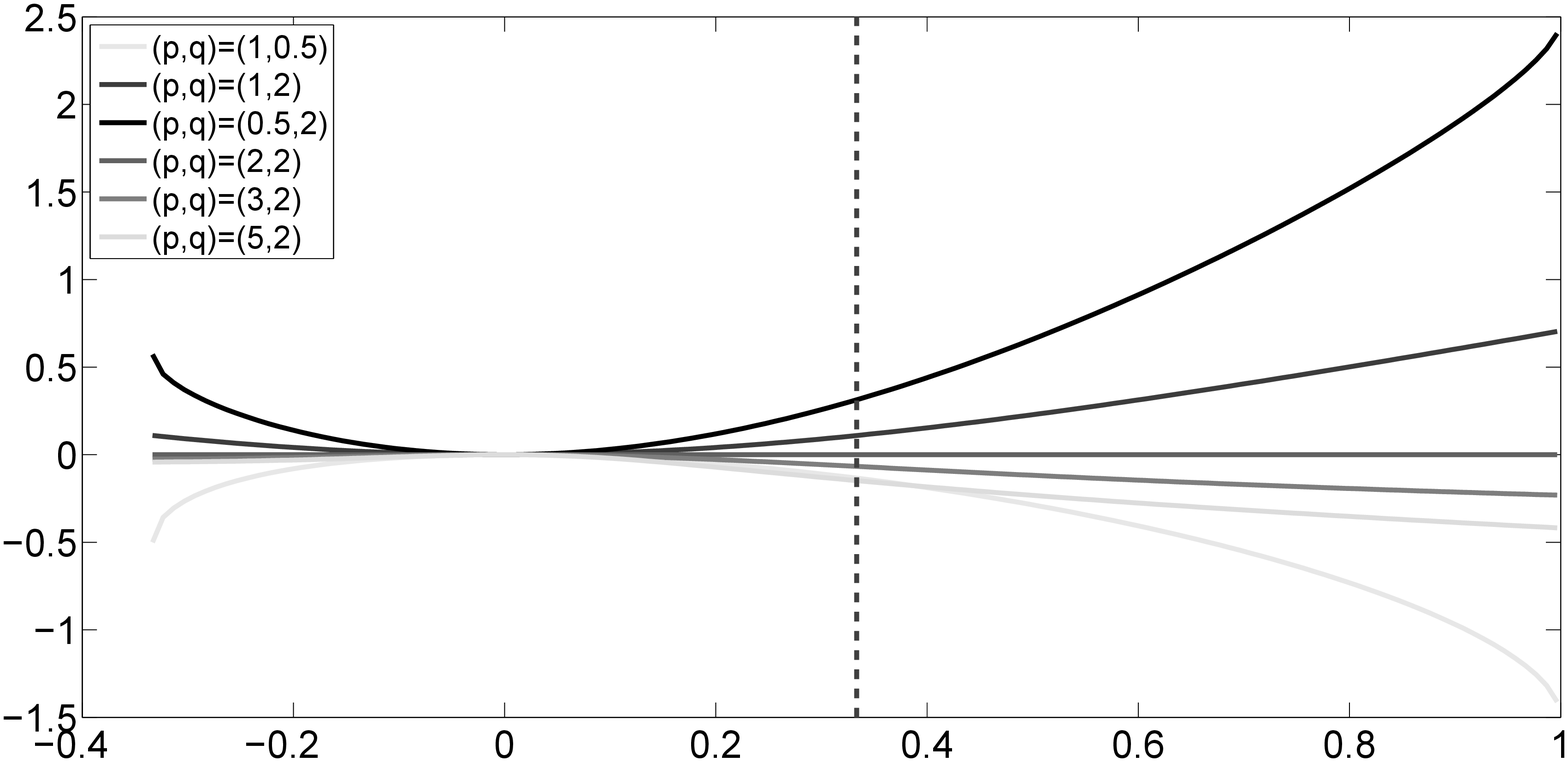, width=15cm}}
\vspace*{8pt}
\caption{The $\Delta \mathcal{M}(p,q)$ (Y-axes) for various pairs $(p,q)$ against $r$ (X-axes), where the dashed line marks $r=1/3$.\label{fig:3}}
\end{figure}

 For the qudit system with $j=3/2$ and with the matrix \eqref{1}, the latter inequality can be rewritten as
\begin{eqnarray}{\left(Tr\left(a_{11}(q)+a_{22}(q)\right)^{\frac{p}{q}}\right)^{\frac{1}{p}}\leq
\left(Tr\left(
                            \begin{array}{cc}
                              Tr a_{11}(p) & Tr a_{12}(p)  \\
                             Tr a_{21}(p)  &  Tra_{22}(p)  \\
                            \end{array}
                          \right)^{\frac{q}{p}}\right)^{\frac{1}{q}}}\,. \label{15}
\end{eqnarray}
Let us denote the  residual of the left and right sides of \eqref{15} as $\Delta \mathcal{M}(p,q)$.
The  $\Delta \mathcal{M}(p,q)$ of the Werner state \eqref{2} is shown in Figures \ref{fig:2} and \ref{fig:3} for various values of parameters $p$ and $q$.

\par For the qutrit state with the density matrix
\begin{eqnarray}\rho={\left(
                                 \begin{array}{ccc}
                                   \rho_{11}& \rho_{12}& \rho_{13}\\
                                   \rho_{21}& \rho_{22}& \rho_{23}\\
                                   \rho_{31}& \rho_{32}& \rho_{33}\\
                                 \end{array}
                               \right),\quad \rho=\rho^{\dagger},\quad Tr\rho=1}\, \label{17}
                               \end{eqnarray}
we can use the results of \cite{Markovich4:2014}. To this end, we
add the matrix (\ref{17}) with zero row and column so that its dimension becomes $4\times4$, i.e.
\begin{eqnarray}\widetilde{\rho}={\left(
                                 \begin{array}{cccc}
                                   \rho_{11}& \rho_{12}& \rho_{13}& 0\\
                                   \rho_{21}& \rho_{22}& \rho_{23}& 0\\
                                   \rho_{31}& \rho_{32}& \rho_{33}& 0\\
                                   0& 0& 0&0\\
                                 \end{array}
                               \right)}\,. \label{11}
                               \end{eqnarray}
Hence, it can be rewritten in the block form as in Sec.~\ref{sec:2}. For example, we get
\begin{eqnarray*}\widetilde{\rho}&=&\left(
                            \begin{array}{cc}
                              a_{11} & a_{12}  \\
                             a_{21}  & a_{22}  \\
                            \end{array}
                          \right),
\end{eqnarray*}
\begin{eqnarray*}a_{11}&=&\left(
                            \begin{array}{cc}
                              \rho_{11} & \rho_{12}  \\
                              \rho_{21}  & \rho_{22}  \\
                            \end{array}
                          \right),\quad a_{12}=\left(
                            \begin{array}{cc}
                              \rho_{13} & 0  \\
                              0  & 0 \\
                            \end{array}
                          \right),\quad
                          a_{21}=\left(
                            \begin{array}{cc}
                              \rho_{31} & \rho_{32}  \\
                              0  & 0  \\
                            \end{array}
                          \right),\quad a_{22}=\left(
                            \begin{array}{cc}
                              \rho_{33} & 0  \\
                              0  & 0 \\
                            \end{array}
                          \right).
\end{eqnarray*}
In these notations, the inequality \eqref{15} is true for the qutrit system. Analogously, any density matrix can be added with zero rows and columns and the Minkowski type inequality \eqref{18} can be used for them.

\section{Conclusion}\label{sec:5}

To conclude we point out the main results of our work. We develop the approach of Ref.~\refcite{Lieb2,Cher} and extend the two-parameter Minkowski type trace inequality to the case of the systems without subsystems. The inequalities are written for the qudit and  qutrit systems. The results are demonstrated on the examples of Werner and $X$-states.


\begin{thebibliography}{0}

\bibitem{Lieb1}E. A. Carlen, E. H. Lieb,  {\it A Minkowski Type Trace Inequality and Strong Subadditivity of Quantum Entropy}, arXiv:math/0701352 (2007).
\bibitem{Lieb2}E. A. Carlen, E. H. Lieb, {\it A Minkowski Type Trace Inequality and Strong Subadditivity of Quantum Entropy II: Convexity and Concavity}, {\it Lett. Math. Phys.} {\bf 83}, 107, (2008), arXiv:0710.4167.
\bibitem{Cher}V. N. Chernega, O. V. Man'ko, V. I. Man'ko, {\it Minkovskii-type inequality for arbitrary density matrix of composite and
noncomposite systems}, (2014) arXiv:1406.5838.
\bibitem{Hardy}
G. H. Hardy, J. E. Littlewood, G. Pólya,   {\it Inequalities}, (Cambridge Mathematical Library (second ed.). Cambridge: Cambridge University Press, 1952).
\bibitem{Chernega}V.N. Chernega, O.V. Man'ko, V.I. Man'ko, {\it Generalized qubit portrait of the qutritstate density matrix}, {\sl J. Russ. Laser Res.}, \sl{34 }, 4, (2013) 383--387.
\bibitem{Chernega:14}V.N. Chernega, O.V. Man'ko, {\it Tomographic and improved subadditivity conditions for two qubits and qudit with j = 3/2}, {\sl J. Russ. Laser Res.}, \sl{35 (1)}, (2014) 27--38 .
\bibitem{OlgaMankoarxiv}V.N. Chernega, O.V. Man'ko, V.I. Man'ko, {\it Subadditivity condition for spin-tomograms and density matrices of arbitrary composite and noncomposite qudit systems}, {\sl arXiv:1403.2233}, (2014).
    \bibitem{Manko:2014}
M.A. Man'ko and V.I. Man'ko (2014), arXiv:1312.6988 \it{Quantum strong subadditivity condition for systems without subsystems}, Phys. Scr., T160, 014030.
\bibitem{Braunstein} S. L. Braunstein, S. Ghosh, S. Severini, {\it Laplacian of a graph as
a density matrix: a basic combinatorial approach to separability of mixed
states}, \sl{Annals of Combinatorics}, {\bf10},  (2006) 291--317.
\bibitem{Bose} S. Bose, A. Casaccino, S.Mancini, S. Severini,
\it{Communication in xyz all-to-all quantum networks with a missing link},
\sl{International Journal of Quantum Information}, \sl{7}, 4, (2009) 713--723.
\bibitem{Werner}R. F. Werner, {\it Quantum states with Einstein-Podolsky-Rosen correlations admitting a hidden-variable model}, {\it Phys. Rev. A}, {\bf40}, (1989) 4277.
\bibitem{Hedemann}S. R. Hedemann, {\it Evidence that All States Are Unitarily Equivalent to X States of the Same Entanglement}, {\sl arXiv:1310.7038}, (2014).
\bibitem{Markovich3:2014}
V.I. Man'ko and L.A. Markovich, \it{Separability and entanglement of the qudit X-state with $j=3/2$}, arXiv:1404.1454, (2014).
\bibitem{Markovich2:2014}V.I. Man'ko and L.A. Markovich, {\it New inequalities for quantum  von Neumann and tomographic mutual information},  {\sl J. Russ. Laser Res.}, \textbf{35(4)},(2014)
\bibitem{Dodonov} V.V. Dodonov, V.I. Man'ko, {\it Positive distribution description for spin states}, {\sl Phys. Lett. A}, \textbf{229}, 335--339 (1997).
\bibitem{OVManko}V.I. Man'ko,  O.V. Man'ko, {\it Spin state tomography}, {\sl JETP}, \textbf{85 (3)}, 430 (1997).
\bibitem{Markovich4:2014}
V.I. Man'ko and L.A. Markovich, \it{ Separability and entanglement of spin $1$ particle}, (submitted in Quantum Information and Computation), arXiv:1406.7118, (2014).


\end{thebibliography}
\end{document}